\documentclass[twocolumn,superscriptaddress,nofootinbib,preprintnumbers,amsmath,amssymb,floatfix]{revtex4-1}

\usepackage{graphicx}
\usepackage{dcolumn}
\usepackage{bm}
\usepackage{xcolor}
\usepackage{subfigure}
\usepackage{hyperref}

\begin{document}

\preprint{}

\title{Elliptic and triangular flow of heavy flavor in heavy-ion collisions}

\author{Marlene Nahrgang}
\affiliation{Department of Physics, Duke University, Durham, North Carolina 27708-0305, USA}

\author{J\"org Aichelin}
\affiliation{SUBATECH, UMR 6457, Universit\'e de Nantes, Ecole des Mines de Nantes,
IN2P3/CNRS. 4 rue Alfred Kastler, 44307 Nantes cedex 3, France }

\author{Steffen Bass}
\affiliation{Department of Physics, Duke University, Durham, North Carolina 27708-0305, USA}

\author{Pol Bernard Gossiaux}
\affiliation{SUBATECH, UMR 6457, Universit\'e de Nantes, Ecole des Mines de Nantes,
IN2P3/CNRS. 4 rue Alfred Kastler, 44307 Nantes cedex 3, France }

\author{Klaus Werner}
\affiliation{SUBATECH, UMR 6457, Universit\'e de Nantes, Ecole des Mines de Nantes,
IN2P3/CNRS. 4 rue Alfred Kastler, 44307 Nantes cedex 3, France }

\date{\today}

\begin{abstract}
We investigate the elliptic and the triangular flow of heavy mesons in ultrarelativistic heavy-ion collisions at RHIC and the LHC. The dynamics of heavy quarks is coupled to the locally thermalized and fluid dynamically evolving quark-gluon plasma. The elliptic flow of $D$ mesons and the centrality dependence measured at the LHC is well reproduced for purely collisional and bremsstrahlung interactions. Due to the event-by-event fluctuating initial conditions from the EPOS2 model, the $D$ meson triangular flow is predicted to be nonzero at $\sqrt{s}=200$~GeV and $\sqrt{s}=2.76$~TeV. We study the centrality dependence and quantify the contributions stemming from flow of the light bulk event and the hadronization process. The flow coefficients as response to the initial eccentricities behave differently for heavy mesons than for light hadrons due to their inertia. Higher-order flow coefficients of heavy flavor become important in order to quantify the degree of thermalization.
\end{abstract}


\maketitle

\section{Introduction}
The evolution of hot and dense QCD matter created in ultrarelativistic heavy-ion collisions is remarkably well described by fluid dynamics. Over the recent years several studies have successfully described $p_T$-spectra and collective flow coefficients measured by RHIC and LHC experiments. Agreement between fluid dynamical calculations and experimental data favors low values for the ratio of shear viscosity over entropy density in the range $\eta/s=0.08-0.24$ \cite{Song:2010mg,Qiu:2011hf,Gale:2012rq}, which indicates the formation of an almost ideal fluid. This success was supported strongly by the precise measurements of higher-order flow coefficients in the light hadron sector \cite{Adams:2004bi,Alver:2006wh,ALICE:2011ab,Adare:2011tg,Abelev:2012di,ATLAS:2012at,CMS2012}. 
While nowadays the equation of state is well constraint by lattice QCD calculations \cite{Borsanyi:2013bia,Bazavov:2014noa}, the initial conditions remain a major source of uncertainty in extracting $\eta/s$ \cite{Song:2010mg}. 
Open questions include many aspects ranging from equilibration times, possible pre-equilibrium dynamics, initial correlations to the treatment of multiple scatterings. In addition, a possible core-corona separation and the hadronization process introduces further uncertainties. Consequently, different setups can describe the data using different optimal combinations of initial conditions and values of $\eta/s$. 

The dynamics of heavy quarks is very different from the light partons forming the bulk of the medium.  Heavy quarks are produced predominantly  in the initial hard scatterings and are not expected to be in equilibrium with the light partons at the formation time of the quark-gluon plasma (QGP). At this initial time $\tau_0$ of fluid dynamics there is thus a clear separation between the collective nature of the bulk and the out-of-equilibrium evolution of the heavy quarks. Light hadron flow builds up as a fluid dynamical response to the initial spatial eccentricities mediated by pressure gradients and is mostly sensitive to the flow in the fluid dynamical medium over the hypersurface at decoupling. 
In order to transfer this bulk flow to the heavy quarks a high interaction rate with the medium constituents is required. Due to the larger masses of charm and bottom quarks, inertia effects limit the efficiency of each interaction process to transfer flow. At early times the temperatures are high and thus the scattering rate of heavy quarks with the medium constituents is large. The efficiency of the early times is, however, balanced by the time needed to develop the flow of the bulk matter, which can then be transformed to the heavy quarks. The experimentally determined light hadron $v_2$ and the $v_2$ of heavy-flavor decay electrons or of $D$ mesons are surprisingly similar \cite{Adams:2004bi,Adler:2005ab,Adare:2011tg,Abelev:2012di,ATLAS:2012at,Chatrchyan:2012ta,Abelev:2014ipa}. The final flow of heavy quarks thus results from integration over the whole evolution time. In addition, heavy-flavor flow receives contributions stemming from energy loss and from the coalescence with light quarks at hadronization.

In this work we demonstrate that besides the well studied elliptic flow also the triangular flow of $D$ mesons is nonvanishing. Investigating the centrality dependence of the flow coefficients we are able to reveal significant differences between the light hadrons, $D$ and $B$ mesons. For this purpose we couple the Monte-Carlo Boltzmann propagation of heavy quarks (MC@sHQ) \cite{Gossiaux:2010yx,Gossiaux:2008jv} to the fluid dynamical evolution of the light bulk sector stemming from EPOS2 initial conditions \cite{Werner:2010aa,Werner:2012xh}: These are obtained from a multiple scattering approach combining pQCD calculations of the hard scattering and Gribov-Regge theory of the soft, phenomenological part of the interaction. Multiple scatterings form parton ladders, which are identified with flux tubes and mapped to the initial fluid dynamical profiles after identifying and subtracting jet components. In the following a $3+1$ dimensional ideal fluid dynamical expansion is performed. Viscosity effects are mimicked by enhancing the initial flux tube radii. 
This EPOS2 version has been applied succesfully to various bulk and jet observables at RHIC and the LHC in A+A collisions \cite{Werner:2010aa,Werner:2012xh}. A recent upgrade, EPOS3, includes a viscous fluid dynamical evolution, based on \cite{Karpenko:2013wva}, and has yielded very good agreement with data from p+p and p+A collisions \cite{Werner:2013ipa, Werner:2013tya}. We plan to couple MC@sHQ to EPOS3 in future work.

The interactions of the heavy quarks with the thermal partons of the fluid are given by either purely collisional scatterings or including gluon bremsstrahlung. The elastic cross sections are calculated within the Born approximation of perturbative QCD including a running coupling $\alpha_s$. The infrared divergence in the $t$-channel is regularized by a self-consistently determined Debye mass $m_D$ \cite{Peshier:2006ah} and thus the gluon propagator is taken as $1/t\to1/(t-\kappa m_D^2(T))$ for all $t$. In the HTL+semihard approach \cite{Gossiaux:2008jv} the parameter $\kappa$ is determined such that the average energy loss is maximally insensitive to the intermediate scale between nominally soft and hard processes. For the incoherent gluon radiation spectra the calculation in \cite{Gunion:1981qs} are extended to finite quark masses \cite{Aichelin:2013mra}. Due to the effect of coherence, i.e. the Landau-Migdal-Pomeranchuk (LPM) effect, the power spectra per unit length are effectively suppressed at sufficiently high $p_T$ \cite{Gossiaux:2012cv}. 
Both interaction mechanisms include uncertainties, which result from various approximations in their formulation, e.g. the neglect of higher-order processes and interference terms, from the effective implementation of the LPM suppression, as well as from uncertainties in the coupling to the background medium such as the interpretation of the equation of state in terms of active degrees of freedom \cite{Nahrgang:2013saa}. We, therefore, 
calibrate the energy loss models by rescaling the cross sections with a global factor $K$. It is chosen such that a reasonable agreement with the $R_{\rm AA}$ data at higher $p_T$ is achieved. For the LHC this gives $K_{\rm c}^{\rm LHC}=1.5$ for the purely collisional and $K_{\rm c+r}^{\rm LHC}=0.8$ for the collisional+radiative(LPM) energy loss mechanism. Given the current range in transverse momentum of the $D$ meson $R_{\rm AA}$ measured at RHIC, this rescaling is less well determined. We assume here that the $K$-factors need to be slightly increased to describe the high-$p_T$ energy loss at RHIC, with $K_{\rm c}^{\rm RHIC}=1.8$ and $K_{\rm c+r}^{\rm RHIC}=1.0$. It is important to note that this allows us to meaningfully compare the heavy-quark flow developed in the two scenarios. Recent results of this model have been presented by some of the authors in \cite{Nahrgang:2013xaa,Nahrgang:2013saa}.

The interaction of heavy flavor in the QGP has attracted a lot of interest over the last years. The $R_{\rm AA}$ and/or the $v_2$ of heavy quarks have been investigated in Langevin-dynamics \cite{Moore:2004tg,vanHees:2004gq,Alberico:2011zy,Cao:2013ita} or within a parton cascade \cite{Uphoff:2011ad}.

\section{QGP: initial state and bulk flow}

\begin{figure}[tb]
 \centering
\includegraphics[width=0.48\textwidth]{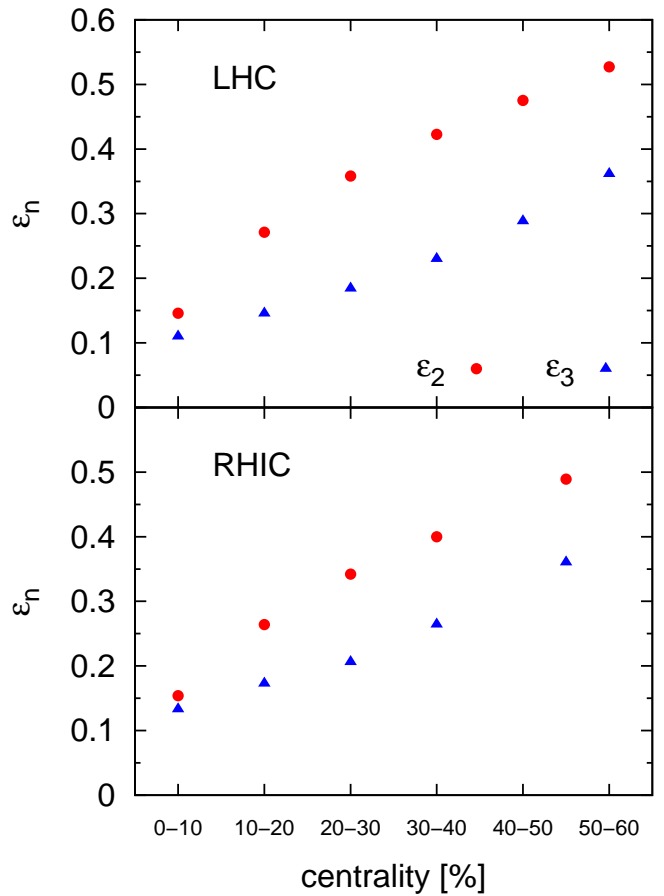}
 \caption{(Color online) The event-averaged initial eccentricities $\epsilon_2$ (red circles) and $\epsilon_3$ (blue triangles) in the EPOS2 model for Pb+Pb collisions at $\sqrt{s}=2.76$~TeV (upper panel) and in Au+Au collisions at $\sqrt{s}=200$~GeV (lower panel) as a function of centrality.}
 \label{fig:eps}
\end{figure}

The fluid dynamical flow of the light bulk sector is the response to the eccentricity in the initial geometry
\begin{equation}
 \epsilon_n=\frac{\sqrt{\langle r^n\cos(n\phi)\rangle^2+\langle r^n\sin(n\phi)\rangle^2}}{\langle r^n\rangle}\, 
\end{equation}
where $\phi$ is the spatial azimuthal angle and $r=\sqrt{x^2+y^2}$ the distance from the center. The average $\langle\cdot\rangle$ is weighted by the local energy density. Similarly the $n$-th order angles of the participant plane\footnote{The term ``participant plane'' is commonly used for the following definition. We would like to point out though, that the initial conditions used here, do not rely on a participant picture.}  can be obtained from the initial state via
 \begin{equation}%
 \psi_n^{\rm PP}=\frac{1}{n}\arctan\frac{\langle r^n\cos(n\phi)\rangle}{\langle r^n\sin(n\phi)\rangle}\, .
 \end{equation}

In Figs.~\ref{fig:eps} we show the event-averaged $\epsilon_2$ and $\epsilon_3$ of the EPOS initial conditions for different centrality classes (based on impact parameter selection) at RHIC and the LHC. While the initial triangularity is exclusively due to fluctuations, the initial ellipticity is dominated by the collision geometry, but due to fluctuations is finite even in the most central collisions. The initial ellipticity $\epsilon_2$ increases faster from central to peripheral collisions than the initial triangularity $\epsilon_3$. Collisions at RHIC and the LHC produce very similar $\epsilon_2$ and $\epsilon_3$.

 \begin{figure*}[tb]
   \centering
  \includegraphics[width=0.95\textwidth]{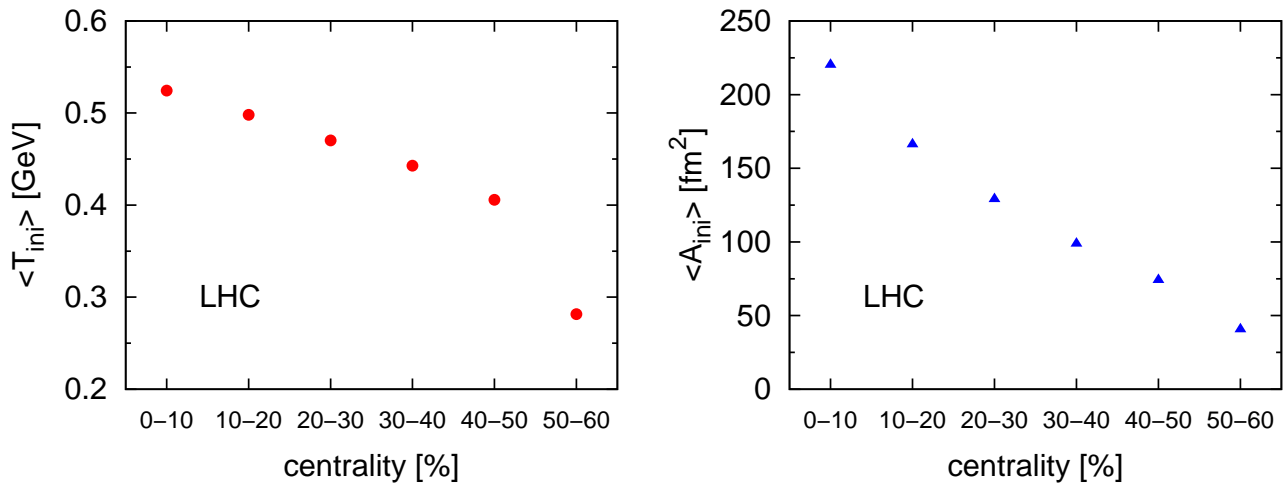}
   \caption{(Color online) The event-averaged initial temperature weighted by the energy density (left) and the average initial overlap area (right), for which $T_{\rm ini}>0.155$~GeV in the x-y plane, in the EPOS2 model of Pb+Pb collisions at $\sqrt{s}=2.76$~TeV as a function of centrality.}
   \label{fig:TVcen}
  \end{figure*}

In extensive theoretical studies of the fluid dynamical response to initial eccentricities it was found that for not too large centralities $v_n$ is proportional to $\epsilon_n$, which even holds fairly well on an event-by-event basis for $n=2$ and to less extent also for $n=3$ \cite{Alver:2010dn,Qin:2010pf,Niemi:2012aj}. The ratio of $v_n/\epsilon_n$ (where both are event-averaged quantities) depends on various aspects, among which viscosity is known to reduce the flow coefficients significantly compared to ideal fluid dynamics and higher-order coefficients are expected to be more sensitive to viscous effects. Details are still under discussion, e.g. if the first-order viscous correction term to $v_n/\epsilon_n$ is $\propto n^2$ \cite{Staig:2011wj,Lacey:2013is,Alver:2010dn} or $\propto n$ \cite{Hatta:2014jva}. The further away from the ideal fluid dynamical behavior the stronger the viscous effects become \cite{Drescher:2007cd}, in general one expects the first-order corrections to be
\begin{equation}
 \frac{v_n}{\epsilon_n}=\left(\frac{v_n}{\epsilon_n}\right)_{\rm ideal}\left(1-{\cal O}(n^mK)\right)\quad m\sim1-2\, ,
\label{eq:vnepsn}
\end{equation}
with the Knudsen number $K$ that relates the relevant microscopic and macroscopic length scales such that $K=0$ is the ideal fluid dynamical limit and $K\to\infty$ is the limit of free streaming.
From Eq.~\ref{eq:vnepsn} one can expect a characteristic decrease of $v_n/\epsilon_n$ with increasing centrality, where the Knudsen number becomes larger.

Already the ideal value of $v_n/\epsilon_n$ does exhibit a centrality dependence, which is due to effects at the edges of the fireball and the freeze-out dynamics \cite{Teaney:2010vd} or due to a core-corona separation as implemented in the EPOS2 model and discussed in \cite{Werner:2010aa}. Furthermore, initially mimicking viscous effects \cite{Werner:2012xh} allows us to meaningfully compare the flow coefficients of light charged hadrons from EPOS2 to data, where good agreement is found.

Experimentally, light hadron flow coefficients have been investigated by all major collaborations in both $p_T$-differential and integrated form \cite{Adams:2004bi,Alver:2006wh,ALICE:2011ab,Adare:2011tg,Abelev:2012di,ATLAS:2012at,CMS2012}. By and large theoretical expectations are met qualitatively while details still need to be settled. An important finding is that the centrality dependence of the triangular flow is smaller than of the elliptic flow. This is due to a competition between the above mentioned phenomena leading to Eq.~\ref{eq:vnepsn} and the centrality dependence of the quantities in Figs.~\ref{fig:TVcen} on the one hand, and the weaker increase of $\epsilon_3$ with centrality as compared to $\epsilon_2$, see Figs.~\ref{fig:eps}, on the other hand.
Also, the fourth- and fifth-order flow coefficients are found to be nonzero for the light charged hadrons with a weak centrality dependence.

The transfer of bulk flow to heavy quarks additionally depends on their interaction strength with the medium constituents. With decreasing temperature the scattering rates decrease significantly and the overall number of scatterings depends on the size of the fireball. Since both, the average temperature and the size of the fireball decrease toward more peripheral collisions, see Fig.~\ref{fig:TVcen}, we expect that the efficiency of transferring bulk flow to the heavy quarks drops quicker as a function of centrality than it does for the light hadrons. In addition, a mass hierarchy should be visible, where the slope of the decrease should be larger for heavier particles with an overall smaller magnitude of the integrated flow coefficients. This is due to the inertia the heavy quarks experience in each interaction with a light parton of the medium.

\section{Heavy-flavor flow}

\begin{figure*}[tb]
 \centering
\includegraphics[width=0.98\textwidth]{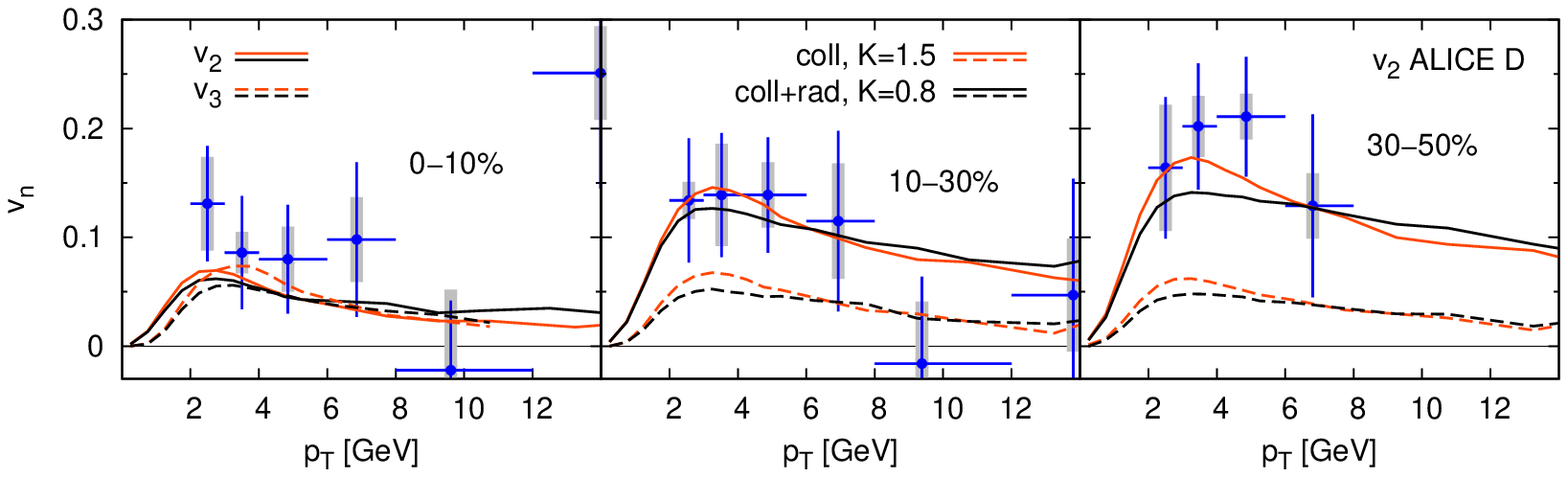}\\
\medskip 

\includegraphics[width=0.98\textwidth]{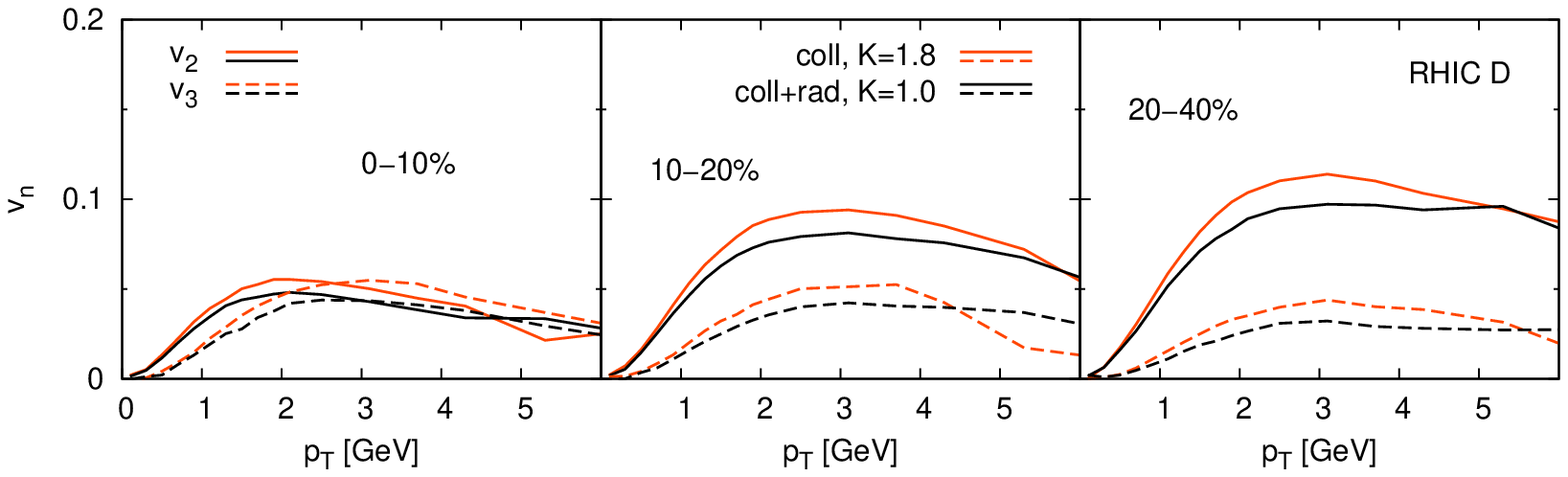}
 \caption{(Color online) $D$ meson elliptic (solid) and triangular (dashed) flow  in Pb+Pb collisions at $\sqrt{s}=2.76$~TeV (upper row) and in Au+Au collisions at $\sqrt{s}=200$~GeV (lower row) for different centralities. Both interaction mechanisms, purely collisional (orange, light) and collisional+radiative(LPM) (black, dark) are shown. Experimental data is taken from \cite{Abelev:2014ipa} for $D^0$. The last two data points for the $30-50$\% most central collisions at the LHC are outside the range of $v_n$ values shown. Simulation results at RHIC suffer from larger statistical fluctuations for $p_T\gtrsim4$~GeV.}
 \label{fig:vn}
\end{figure*}

It has been shown that the flow coefficients
 \begin{equation}
  v_n^{\rm EP}=\frac{\int{\rm d}\phi\cos\left[n(\phi-\psi_n^{\rm EP})\right]\frac{{\rm d}N}{{\rm d}y{\rm d}\phi}}{\int{\rm d}\phi\frac{{\rm d}N}{{\rm d}y{\rm d}\phi}}\, ,
 \end{equation}
taken as the Fourier coefficients of the single-particle azimuthal distribution with respect to the event-plane angle $\psi_n^{\rm EP}=(1/n)\arctan(\langle p_T\sin(n\phi)\rangle/\langle p_T\cos(n\phi)\rangle)$, where $\phi$ is the azimuthal angle of the transverse momentum of the measured particles, corresponds very well to the flow coefficients $v_n^{\rm PP}$ obtained from correlating the single particles with the initial participant plane \cite{Holopainen:2010gz}.
The experimental data from the ALICE collaboration \cite{Abelev:2014ipa} shows that the $D$ meson elliptic flow obtained from the event-plane method is consistent with that obtained from the scalar product and the two-particle cumulant method. We follow the assumption that the initial participant plane is a good approximation for the final event plane and that our results can thus be directly compared to the experimental data. It will of course be a task of future investigation what the quantitative uncertainties due to this different procedure are, once the final hadronic interactions are included in the model.

In order to optimize computational resources in the full evolution scenarios we evolve $N^{\rm HQ}=10^4$ heavy-quark events per fluid dynamical event and have checked that the results do not change significantly when one further decreases $N^{\rm HQ}$ toward a real event-by-event calculation.

\begin{figure*}[tb]
 \centering
\includegraphics[width=0.95\textwidth]{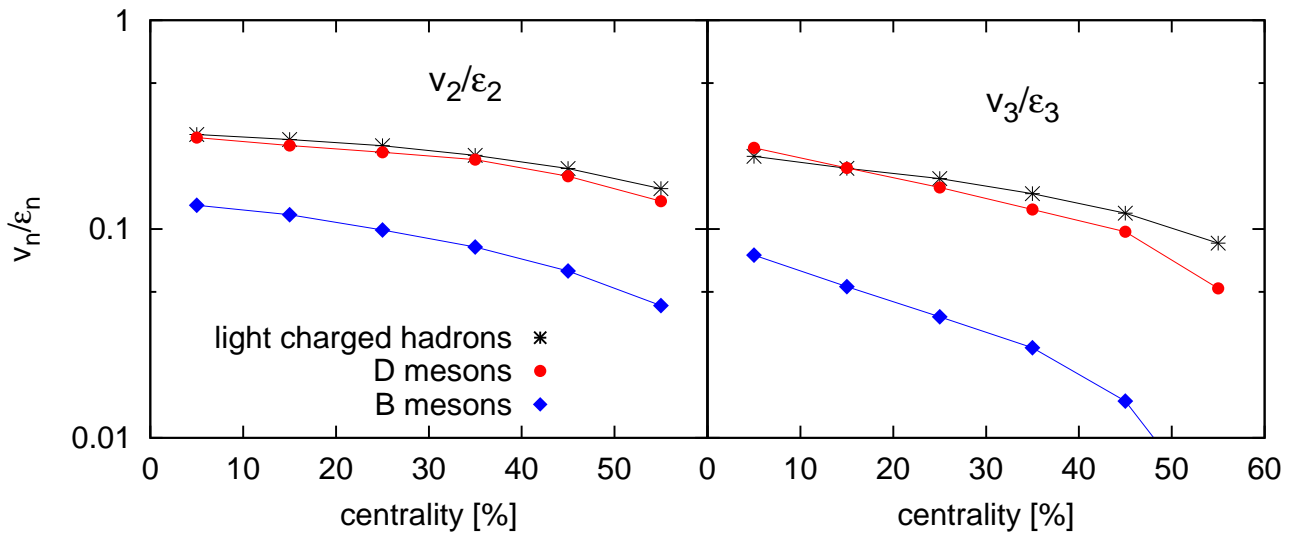}

 \caption{(Color online) The centrality dependence of the ratios $v_n/\epsilon_n$ for $n=2, 3$ in the  collisional+radiative(LPM) energy loss model at $\sqrt{s}=2.76$~TeV collision energies for the light charged hadron, the $D$ and $B$ meson flow. The bulk flow is obtained as $v_n(2,|\Delta\eta|>1)$ from the full EPOS2 model \cite{Werner:2012xh}.}
 \label{fig:vnepsn}
\end{figure*}

Fig.~\ref{fig:vn} (upper row) shows the $p_T$-dependence of the elliptic and triangular flow of $D$ mesons for the three centrality classes $0-10$\%,  $10-30$\% and  $30-50$\% for which $v_2$ data from ALICE is available. For RHIC energies we choose the $0-10$\%,  $10-20$\% and  $20-40$\% most central events, Fig.~\ref{fig:vn} (lower row). For each of these cases $v_{2,3}$ are calculated for the purely collisional and the collisional+radiative(LPM) energy loss mechanism. 
Both interaction mechanisms describe well the $D$ meson $v_2$ at the LHC in the given centralities, if one expects that the hadronic stage can also contribute to the final $D$ meson $v_2$ \cite{He:2011yi,He:2012df,Cao:2014dja,Ozvenchuk:2014rpa}. At lower $p_T\lesssim 5$~GeV the purely collisional interaction suffers less from phase-space limitations than the radiative contributions. It thus yields a larger $v_{2,3}$ than the scenario including bremsstrahlung and is therefore closer to the central data points. For $p_T$ between $2-4$~GeV this enhancement is $10-20$\% for the $v_2$ in the most peripheral centrality class at the LHC and RHIC and $20-30$\% (LHC) and $30-40$\% (RHIC) for $v_3$. 
At larger momentum this picture is reversed since here the path length difference is the main contribution to the second Fourier coefficient. The origin of this effect is an interplay between an initial $p_T$ spectrum, which falls off quickly as a function of $p_T$, like the proton-proton reference spectrum used here, and a model, where the energy loss increases with the $p_T$ of the heavy quark, like the models investigated here. A larger $v_2$ in the collisional+radiative(LPM) scenario is then expected because the average energy loss increases faster as a function of $p_T$ than in the purely collisional case. Toward more central collisions both, the flow of the medium as well as the path length differences, diminish and so does the heavy quark $v_2$ over the entire $p_T$ range.

The triangular flow of $D$ mesons shows very little centrality dependence. As mentioned earlier the measured triangular flow of charged particles increases only very weakly with centrality \cite{Abelev:2012di,ATLAS:2012at} and the same is seen in theoretical studies of anisotropic flow of thermal photons \cite{Shen:2013cca}. While the centrality dependence of the bulk flow is weak the larger and hotter medium produced in more central collisions is more effective in transferring this bulk flow to the heavy quarks. As a consequence the triangular flow of $D$ mesons does not show any dependence on centrality. Toward larger centralities we find less coupling of the heavy quarks to the bulk flow.

By comparing the $D$ meson flow developed at RHIC and at the LHC, we find very similar trends but overall smaller peak values at RHIC, which again is due to smaller temperatures and initial volumes produced at the lower energies. The bulk flow might be very similar at both energies due to similar initial eccentricities and values of viscosity, whereas the heavy-flavor flow is expected to be different.

\begin{figure*}[tb]
 \centering
\includegraphics[width=0.48\textwidth]{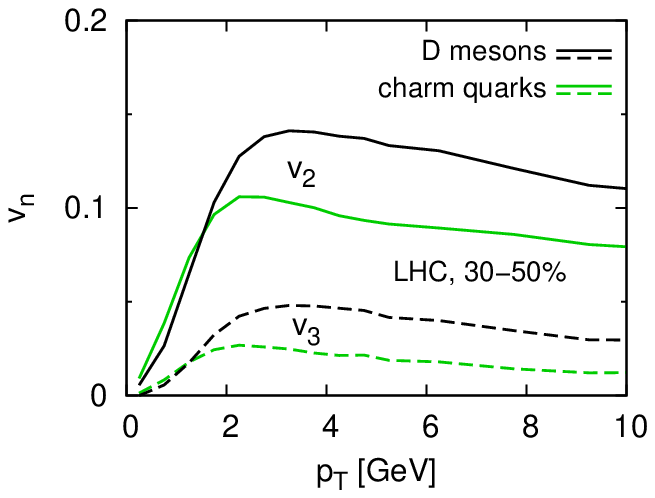}
\includegraphics[width=0.48\textwidth]{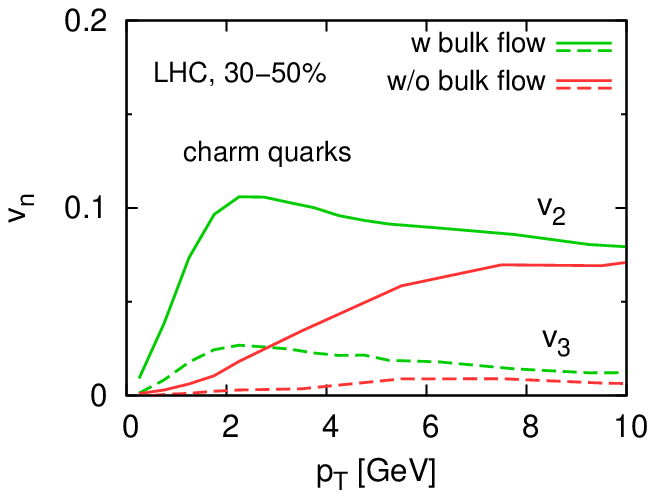}
 \caption{(Color online) The contribution of hadronization to the $D$ meson elliptic (solid) and triangular (dashed) flow (left plot) and the contribution of the bulk flow to the charm quark elliptic (solid) and triangular (dashed) flow (right plot) for $30-50$\% most central Pb+Pb collisions at $\sqrt{s}=2.76$~TeV.}
 \label{fig:flow}
\end{figure*}

We now turn to investigate the centrality dependence of the heavy-flavor flow further by plotting the integrated $v_n/\epsilon_n$, as a function of the centrality in Fig.~\ref{fig:vnepsn}. The $\epsilon_n$ are the same average quantities as in Fig.~\ref{fig:eps} and we concentrate on the collisional+radiative(LPM) energy loss model at $\sqrt{s}=2.76$~TeV. Under the assumption that $v_n\propto\epsilon_n$ (which holds for small and intermediate centralities) the plotted quantity can be identified with the efficiency of the medium to transform an initial geometry into an anisotropy in momentum space. By comparing the $D$ mesons flow to the flow of the light charged hadrons from the bulk and the heavy $B$ mesons we can make the following observations. For all particles we see that the efficiency of the system to respond to the initial geometry decreases toward more peripheral collisions and a mass hierarchy can be observed in the slopes of this decrease. 
For $v_2/\epsilon_2$, $D$ mesons and light charged hadrons show a very similar behavior in both the magnitude and the slope, which as such would imply that the overall efficiency of transferring an initial ellipticity to bulk flow and to flow of the charm quarks is of the same order suggesting a perfect coupling of the charm quarks to the bulk. We can see, however, that this does not hold for the third-order Fourier coefficient of the flow, where although being of the same magnitude (within expected errors)  in the central collisions, the ratio $v_3/\epsilon_3$ falls off more quickly for $D$ mesons than for the bulk flow toward more peripheral collisions. In the present study, all $v_{n>3}$ of the heavy mesons were found to be very small in central collisions and zero within the statistical fluctuations for larger centralities. For $B$ mesons, the flow is smaller in magnitude and by the steeper decrease one can see a more rapid decoupling from the bulk medium that can be understood as an effect of inertia due to the mass hierarchy. Again, this is more pronounced in the third-order Fourier coefficient.
Fig.~\ref{fig:vnepsn} also illustrates nicely that the incomplete coupling due to reduced temperatures and volumes balances an increase in the initial triangularity, such that we do not observe a centrality dependence of $D$ meson $v_3(p_T)$ in Fig.~\ref{fig:vn}.
The present analysis uses a two-particle correlation method with a gap in $|\Delta\eta|<1$ for the determination of the flow coefficients of the bulk, as has been done in the full EPOS2 model \cite{Werner:2012xh}. The results agree well with experimental data in \cite{ALICE:2011ab}. We leave it for future work in a fully coupled model including hadronic rescatterings of the heavy mesons and dynamical viscous corrections to quantify the corrections stemming from using the reconstructed event plane instead of the initial participant plane for determining the heavy-meson flow.

In the following we study several contributions to the heavy-flavor flow by concentrating on the collisional+radiative(LPM) energy loss model in peripheral collisions at $\sqrt{s}=2.76$~TeV. In the left plot of Fig.~\ref{fig:flow} the effect of hadronization on the heavy-flavor flow is shown. It compares the charm quark flow obtained at the hadronization hypersurface to the $D$ meson flow after hadronization. The default hadronization procedure in our model includes coalescence, which happens predominantly at lower $p_T$, and fragmentation, which dominates at high $p_T$. The heavy quarks at low $p_T$ pick up additional $p_T$ and flow from the bulk medium, which leads to an increase of both the $v_2$ and the $v_3$ of $D$ mesons compared to the one of charm quarks above $p_T\gtrsim1.5$~GeV.

The charm quark $v_2$ is about $75$\% of the $D$ meson $v_2$ for $p_T\gtrsim2$~GeV, while the charm quark $v_3$ accounts for only $50$\% of the $D$ meson $v_3$.

In theoretical studies we can artificially switch off the bulk flow by assuming that the local rest frame of the fluid is the same as the laboratory frame. This procedure is of course only a first approximation to a scenario without bulk flow as the temperature field is still taken from an evolution that includes bulk flow, yet it gives an idea of how much of the heavy-flavor flow stems from the path length difference due to the initial eccentricity.
In accordance with previous studies \cite{Gossiaux:2012ea,Cao:2012jt} we find that around $p_T\sim2$~GeV both the $v_2$ and the $v_3$ of charm quarks are almost entirely due to the bulk flow of the medium. At $p_T\approx4$~GeV the charm quark $v_2$ originating from path length differences is $\sim50$\% of the charm quark $v_2$ produced in a medium with bulk flow. This picture is slightly different for the triangular flow $v_3$. Path length differences seem to be smaller in trianguarly shaped event geometries and the corresponding angular sectors are smaller, which diminishes the importance of this contribution to the flow. Up to $p_T\sim4$~GeV we find that the charm quark $v_3$ is built up almost exclusively from the bulk flow of the medium, which makes it an excellent probe of the dynamics and interactions of charm quarks in the quark-gluon plasma.

 \section{Conclusions}\label{sec:conclusions}
We have investigated the development of heavy-quark flow in a fluid dynamical description of the bulk medium. Comparing to the experimental data for the  $D$ meson $v_2$ in three different centrality classes, we found a good overall agreement for both energy loss models, purely collisional and collisional+radiative(LPM). The trend shows that the elliptic flow of $D$ mesons developed in the partonic phase is not quite sufficient to describe all the data well. It would be worth to further study the influence of hadronic interactions of the $D$ mesons. 
Due to the fluctuating initial conditions and the  interactions of the charm quarks with the medium constituents we can present the first predictions of a finite triangular flow of $D$ mesons in heavy-ion collisions at RHIC and LHC energies. For low transverse momentum we find that both the elliptic flow and the triangular flow are larger in a purely collisional energy loss model. The relative enhancement of the flow in a purely collisional scenario over that in a scenario including gluon bremsstrahlung is larger for the triangular flow than for the elliptic flow, and larger at RHIC than at the LHC. 
We did not observe a centrality dependence of the triangular flow like the one observed for elliptic flow. This can be explained by a cancellation of the bulk flow which increases weakly with centrality and the hotter and larger QGP medium created in central collisions. Higher temperatures and larger volumes are more effective in transferring the bulk flow to the heavy quarks via a larger scattering rate and an overall longer time available for interactions. The centrality dependence of $v_n/\epsilon_n$ supports this claim. In addition, it shows that higher-order Fourier coefficients are more likely to show an incomplete coupling of the heavy quarks to the bulk medium as well as the expected mass hierarchy.

Approximately $50$\% of the triangular flow of $D$ mesons above $p_T\gtrsim2$~GeV originates from the flow of the charm quarks. 
We could further show that the contributions from path length differences are suppressed for the triangular flow as compared to the elliptic flow and at low $p_T\sim2$~GeV both flow coefficients are almost entirely due to the bulk flow in medium.

The study of heavy-flavor flow is an ideal tool to learn about the heavy-quark dynamics in the QGP, the basic interactions and the hadronization process. The present study is the first investigation of the potential of higher-order flow coefficients of heavy quarks. Due to many rescatterings in the medium and the hadronization process heavy quarks can probe different aspects than for example the study of the anisotropic flow of thermal photons. For a more complete picture it would be enlightening to study both of these medium probes in the same framework. 

More studies about the relation of the flow of the bulk medium and the heavy-flavor flow are under way. With the upcoming version of viscous fluid dynamics from EPOS3 initial conditions and the inclusion of a hadronic stage, this shall open new opportunities to perform even more realistic comparisons with experimental data and to study the particular role heavy quarks play in the investigation of the QGP.

\section*{Acknowledgments}
M.N. was supported by a fellowship within the Postdoc-Program of the German Academic Exchange Service (DAAD). This work was supported by the U.S. department of Energy under grant DE-FG02-05ER41367, by the TOGETHER project R\'egion Pays de la Loire and I3-HP.

\bibliography{biblio_nahrgang}

\end{document}